# Interplay of pulse duration, peak intensity, and particle size in laser-driven electron emission from silica nanospheres

**Jeffrey A. Powell**[1,2,3], **Adam M. Summers**[1], **Qingcao Liu**[4], **Seyyed Javad Robatjazi**[1], **Philipp Rupp**[4], **Johannes Stierle**[4], **Carlos Trallero-Herrero**[1,2], **Matthias F. Kling**[4,5], and **Artem Rudenko**[1*]

[1]*J.R. Macdonald Laboratory, Department of Physics, Kansas State University, Manhattan, Kansas 66506, USA*
[2]*Department of Physics, University of Connecticut, Storrs, Connecticut 06269, USA*
[3]*INRS, Énergie, Matériaux et Télécommunications, 1650 Bld. Lionel Boulet, Varennes, Québec, J3X 1S2, Canada*
[4]*Physics Department, Ludwig-Maximilians-Universität Munich, D-85748 Garching, Germany*
[5]*Max Planck Institute of Quantum Optics, D-85748 Garching, Germany*
*\*rudenko@phys.ksu.edu*

**Abstract:** We present the results of a systematic study of photoelectron emission from gas-phase dielectric nanoparticles ($SiO_2$) irradiated by intense 25 fs, 780 nm linearly polarized laser pulses as a function of particle size (20 nm to 750 nm in diameter) and laser intensity. We also introduce an experimental technique to reduce the effects of focal volume averaging. The highest photoelectron energies show a strong size dependence, increasing by a factor of six over the range of particles sizes studied at a fixed intensity. For smaller particle sizes (up to 200 nm), our findings agree well with earlier results obtained with few-cycle, ~4 fs pulses. For large nanoparticles, which exhibit stronger near-field localization due to field-propagation effects, we observe the emission of much more energetic electrons, reaching energies up to ~200 times the ponderomotive energy. This strong deviation in maximum photoelectron energy is attributed to the increase in ionization and charge interaction for many-cycle pulses at similar intensities.

## 1. Introduction

Isolated nanosystems such as clusters [1–4], nanoparticles [5–8], nanowires [9,10] and nanotips [11,12] represent perfect test grounds for studies of light-matter interactions with increasing levels of complexity. The ability to precisely synthesize these systems with desired shapes, sizes, and composition [13,14] allows for a rich set of parameters, which can be utilized to study various aspects of their response to an external optical field [15,16]. Photoelectron emission represents one of the most fundamental types of such responses. Recently, considerable effort has been spent to understand the electron dynamics in individual nanoparticles driven by intense, femtosecond pulses, largely motivated by the vision of controlling electronic motion at the nanoscale [6,12,17–22]. The results reveal many important features highlighting the transition from atomic to bulk matter responses to intense fields. In particular, these spectra carry signatures of laser-induced near-field enhancements [6] and dynamical many-particle charge interactions [19], processes not present in atomic or molecular systems.

In this work, we specifically focus on photoelectron emission from spherical $SiO_2$ nanoparticles irradiated by 780 nm, 25 fs (~10 cycles) laser pulses as a function of particle size and laser intensity. The local intensity is a key factor in determining the mechanisms and subtleties in the photoelectron emission. While the physical parameters of the nanoparticle target can be well-defined, the effective laser intensity experienced by an individual nanoparticle is not precisely known due to the inherent spatial distribution of the particle beam in the laser focus. Nanoparticles in the wings of the laser focus necessarily experience a lower intensity than those exposed to the center, peak intensity. Consequently, the measured observables, e.g., photoelectron energy and angular distributions, effectively average over the spatial intensity distribution of the laser field, which significantly complicates their interpretation and comparison with theory.

Here, we present an experimental technique aimed at studying both intensity- and size-dependent photoelectron emission patterns from gas-phase nanoparticles. To avoid focal volume averaging, we employ an approach conceptionally similar to the sorting technique proposed by Gorkhover *et al.* [23]. It is based, however, in our case on using the number of emitted photoelectrons per laser shot for a given nanoparticle size as a relative measure of the local laser intensity (i.e. the position of the particle within the laser focus). By sorting results according to this observable, accurate energy- and angle-resolved photoelectron spectra corresponding to a narrow incident laser intensity range within the laser focus are obtained. We illustrate this technique using the strong-field electron emission from $SiO_2$ nanospheres.

We show that the photoelectron cutoff energy is dominated by the nanoparticles ionized at the peak intensity of the laser pulse (i.e. in the center of the focal profile). The cutoff energy shows a strong size dependence, increasing by a factor of six over the range of particle sizes studied at a fixed intensity. We compare our results to earlier data obtained using few-cycle pulses [19]. This comparison is performed by normalizing the measured cutoff energy to the so-called ponderomotive potential $U_p = \frac{e^2 I}{2c\epsilon_0 m_e \omega^2}$ (the cycle-averaged quiver energy of a free electron in an electromagnetic field) to account for the somewhat different intensities and wavelengths used in both measurements. (Here, $\omega$ is the frequency of the oscillating electric field, $I$ - its intensity, $e$ and $m_e$ – electron charge and mass, respectively). For small particle sizes (diameter $d \leq 200$ nm) our results agree well with those for few-cycle pulses [19]. However, for larger nanoparticles, for which the dimensionless Mie size parameter $\rho = \pi d / \lambda$ (where $\lambda$ is the incident light wavelength) starts to exceed unity, we observe the emission of significantly more energetic electrons. These photoelectron energies reach up to 200 $U_p$ for the largest particles studied. In line with earlier theoretical considerations [24], we qualitatively attribute this effect to the stronger near-field localization for larger particles and simultaneously increased charge creation and interaction for longer pulses.

## 2. Photoelectron emission from nanoscale particles

Detailed information on the interaction dynamics of a femtosecond laser pulse with a nanoparticle can be imprinted in the properties of the emitted photoelectrons. At sufficiently high intensities, multi-photon or tunnel ionization of the nanoparticle occurs. In the atomic case, an emitted photoelectron propagates in the continuum driven by the oscillating laser field where it can gain additional energy. In a simple, semi-classical picture for a linearly polarized field, a "direct" electron (i.e., the electron which does not interact with its parent ion after ionization) can gain up to $2U_p$ of kinetic energy [25]. Photoelectrons driven back to the parent ion can gain larger energies, reaching values up to ~10 $U_p$ (often called "the photoelectron cutoff") for elastically backscattered electrons [25,26].

Analogous to the atomic case, photoelectrons emitted from nanoparticles can be driven back by the oscillating laser field and interact with the particle surface. However, as a nanoparticle is made up of many thousands to millions of individual atoms, many-atom bulk properties significantly alter the photoelectron rescattering processes (elastic or inelastic). The electron propagates within an effective field consisting of the driving laser field, induced near-field of the particle, the "trapping potential" of the left-behind surface ions, and the field created by the interaction with other photoelectrons [17,19,24,27]. Each of these contributions can affect the trajectories and momenta of the freed electrons. The spatial distribution of the induced near-field depends strongly on the size of the system, described by the Mie size parameter $\rho$ [17,19]. The near-field resembles a dipole-like distribution when $\rho \ll 1$, mainly consisting of radial field components (with respect to the laser polarization) resulting in a predominantly radial acceleration for the scattered electrons. However, when $\rho \geq 1$, the dipole distribution begins to break down and higher-order terms significantly contribute, resulting in a shift of the maximum near-field enhancement towards the propagation direction. Previous theoretical work has shown that charge creation and interaction is enhanced in this regime and contributes to the acceleration [24].

Thus, the photoelectron emission properties, in particular, the cutoff energy, can be expected to sensitively depend on both nanoparticle properties (composition, shape, size) and laser pulse characteristics (wavelength, peak intensity, pulse duration). Understanding of how the interplay of these factors shapes the photoelectron emission pattern and determines the highest electron kinetic energy is the main goal of this work.

## 3. Experimental apparatus

Our experiments made use of intense 25 fs, 780 nm laser pulses produced with a chirped pulse amplification (CPA) Ti:Sapphire "PULSAR" laser system running at 10 kHz [28]. Pulse energies for this experiment ranged from 40 µJ to 80 µJ. These laser pulses were focused onto a continuous nanoparticle beam inside a velocity map imaging (VMI) spectrometer. The overall configuration of the experiment is similar to the setup described in [29]. The aerosol-based nanoparticle source delivered single, isolated gas-phase nanoparticles in vacuum. Briefly, a suspension of silica ($SiO_2$) nanoparticles in water was aerosolized and a solid-state membrane dryer selectively removed the solvent (water) from the carrier gas ($N_2$). An aerodynamic lens was used to focus the gas-phase nanoparticles to increase the beam density while a three-stage differential pumping system removes excess carrier gas in vacuum [30,31]. At the 10 kHz repetition rate used, the propagation velocity of the nanoparticle beam was large enough to guarantee a fresh nanoparticle target for each laser shot. Spherical silica nanoparticle samples (nanoComposix, Inc.) ranging from 20 nm to 750 nm (silanol surface coating) were custom ordered specifically for their narrow size distribution (<10%), solvent choice and overall purity. The initial number concentration of each sample was experimentally determined to minimize the probability of more than one nanoparticle per aerosol droplet (dimer formation) while also ensuring less than one particle in the interaction volume per laser shot.

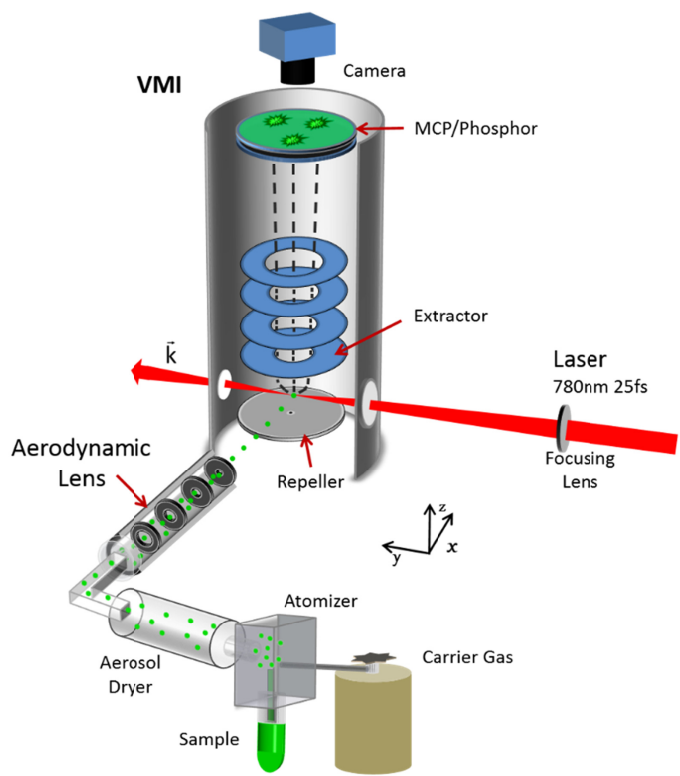

Fig. 1. High energy velocity map imaging (VMI) spectrometer coupled to nanoparticle source. Single, isolated gas-phase nanoparticles are focused and injected into vacuum to interact with a 780 nm, 25 fs, 10kHz laser source. Emitted electrons are focused onto the MCP/phosphor assembly where a single-shot camera records the electron spectra for each laser shot.

A thick-lens, high-energy VMI spectrometer, [32] shown in figure 1, resolved the angular and momentum distribution of all emitted photoelectrons. The spectrometer was able to detect electrons up to 240 eV electron energy and was operated in counting mode. The use of a single-shot camera coupled with a real-time hit finding routine allows for the electron spectra to be captured for each laser shot. Since for the largest nanoparticles used in this work, the electron emission pattern is not cylindrically symmetric with respect to the laser polarization, a systematic inversion of the VMI images yielding full 3D electron momentum distributions for the whole data set was not feasible. As we were particularly interested in the most energetic (cutoff) photoelectrons, for which the upper energy boundaries of the full 3D momentum sphere and of the VMI image scaled in units of momentum (i.e., the 2D projection of the full momentum sphere onto the detector plane) are essentially the same, we restrict ourselves to the non-inverted data throughout the paper.

The peak laser intensity was determined by analyzing the above-threshold ionization (ATI) electron energy distribution of atomic Xe from an effusive jet. We used the ponderomotive shift of the Xe ATI comb, measured as a function of laser pulse energy, to derive the ponderomotive energy and, thus, the peak laser intensity [33].

## 4. Near-single intensity photoelectron imaging

As discussed in the introduction, the nanoparticles in gas-phase experiments typically experience a broad range of laser intensities due to the spatial intensity distribution across the laser focus. The measured observables, e.g., photoelectron energy and angular distributions, are effectively averaged over this intensity distribution of the interaction. For photoionization experiments with isolated atoms and molecules in a multiphoton or tunneling regime, this issue is often somewhat less critical because of the highly non-linear dependence of the ionization probability (and, thus, of the photoelectron yield) on the intensity of the laser, such that the vast majority of the events contributing to the observed spectra originate from a (small) volume corresponding to the peak intensity value. For a nanoscale system, where the number of emitted photoelectrons as a function of the laser intensity is nearly linear [6,24], the resulting spectra directly reflect the spatial profile of the laser focus, heavily favoring the regions with lower intensities which have larger focal volumes.

Here, our "intensity binning" technique allows for the study of intensity and size dependent photoelectron emission patterns from gas-phase nanoparticles, while to a large extent avoiding focal volume averaging. The method uses the number of emitted photoelectrons per laser shot for a given nanoparticle size as a relative measure of the local laser intensity and, thus, the position of the nanoparticle within the laser focus. To achieve this, a histogram (figure 2(a)) showing the number of detected electrons per laser shot is constructed for a measurement performed under a particular set of conditions (fixed peak laser intensity and nanoparticle size). When compared to the background scan (black), the deviation at larger number of electrons per laser shot shows the contribution from the nanoparticles above the background level. A series of bins (grey boxes) are positioned on the histogram, selecting only the laser shots containing this range of electrons for further analysis. The extended width of the histogram in figure 2(a) is indicative of the nanoparticle beam sampling the entire laser focus (focal volume averaging). The inset cross-sectional schematic of a Gaussian focus shows how the different 'bins' map to distinct focal positions, with the focus center containing the peak intensity and the smallest volume.

Figure 2(b) portrays a typical histogram from 120 nm $SiO_2$ nanoparticles as a function of peak laser intensity from $8.8 \times 10^{12}$ W/cm$^2$ to $1.8 \times 10^{13}$ W/cm$^2$. The photoelectron yield increases monotonically with laser intensity as does the cutoff energy. As the intensity grows, the focal volume where ionization occurs also increases. The highest energy photoelectrons originate from ionization events that occur in the central, peak intensity region of the laser focus.

Figure 2(c) shows the full photoelectron angular distribution from the summation of all the laser shots contributing to a specific intensity bin as chosen in figure 2(a). These are 2D, non-inverted VMI images scaled in units of electron momentum. As the number of electrons per laser shot increases, the photoelectron momentum and angular distributions clearly changes. Each bin illustrates the contributions from a particular intensity range within the focus, with distinct differences highlighted between the smallest and largest bin. The full, or integrated, VMI image shows all photoelectron contributions above the background threshold and thus is subjected to this volumetric weighting. This binning technique allows for near-single intensity photoelectron spectra to be obtained and focal volume averaging to be minimized. To analyze the photoelectron emission patterns more quantitatively as a function of bin selection, the radial distribution of each binned 2D VMI image, rescaled to units of energy, is shown in figure 2(d). As mentioned in section 3, the VMI images for the largest nanoparticle sizes are not cylindrically symmetric with respect to the polarization and, therefore, the entire data set was not inverted. However, the upper energy boundaries of the 2D projection versus the full 3D momentum sphere are effectively the same.

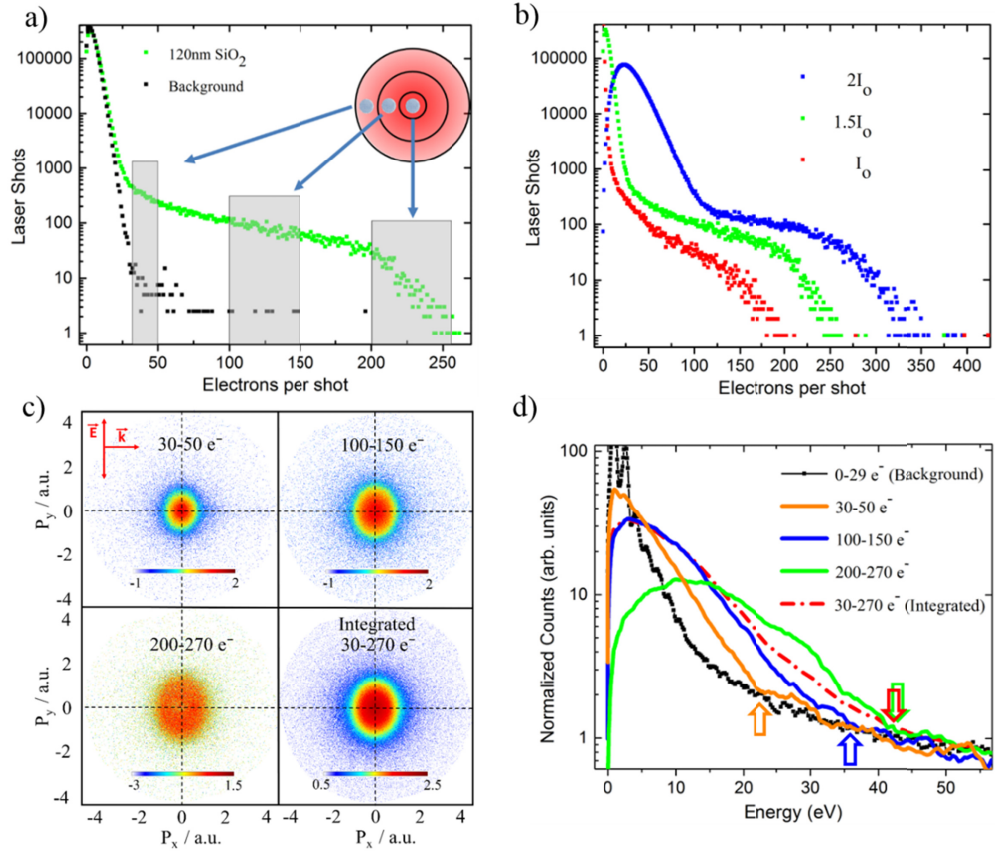

Fig. 2. **a)** Histogram showing the number of laser shots occurring with a given number of photoelectrons detected for 120 nm $SiO_2$ at $1.3 \times 10^{13}$ W/cm$^2$. The background (black) data was measured under identical conditions but without nanoparticles present. A representative transverse slice of a Gaussian focus (inset) depicts the different contributions as a function of focal position. **b)** Similar histogram for the same particle size but multiple peak laser intensities ($I_o = 8.8 \times 10^{12}$ W/cm$^2$). **c)** VMI momentum images (rescaled to atomic units (a.u.) of momentum) from the corresponding 'bins' of the histogram chosen in (a). Each image is the summation of the electron spectra from all laser shots within each 'bin'. The integrated image includes all laser shots above the background level. Color bar in log scale. **d)** Radial distribution of 2D VMI images shown in (c) scaled to units of energy. Only the electrons emitted in a 30° full-opening angle slice along the polarization direction are included. Each 'bin' corresponds to a different sampling of the laser focal volume by using the number of electrons per shot as a guide. The colored arrows indicate the location of the determined cutoff (i.e., the photoelectron energy for which nanoparticle electron spectra start to coincide with the background) for that particular bin. The dash-dotted red line depicts the rescaled radial distribution from all laser shots above the background and shows how a relatively good electron cutoff energy can be determined with a coarse 'binning' procedure, granted this includes the contributions from the largest number of electrons per shot.

The radial distribution of each bin, including the integrated spectra, was vertically shifted until their high energy tails overlapped with that of the background. As the background distribution originates from the laser shots that do not contain any nanoparticles in the laser focus, it represents a good reference to compare to the photoelectrons emerging from a nanoparticle. The cutoff energy was defined as the photoelectron energy for which the nanoparticle signal statistically falls into the background. Arrows in figure 2(d) show the cutoff values determined for each image presented in figure 2(c). It can be clearly seen that each bin with a larger number of electrons (i.e. resulting from a higher local laser intensity) also has a larger cutoff energy. An additional noteworthy observation is that the integrated spectrum (dashed red line) essentially coincides with the distribution for the largest bin (200-270 detected electrons, green line in figure 2(d)) at high electron energies, indicating that the cutoff energies for the largest bin and the integrated spectra are essentially the same. This confirms that the highest energy photoelectrons are emitted from the nanoparticles experiencing the largest intensity, necessitating the cutoffs for the largest bin and for the integrated distribution in an "ideal" experiment to be equivalent. We used the integrated spectra when determining the photoelectron cutoff for our subsequent data.

## 5. Size Dependent Photoelectron Emission

Spherical silica nanoparticles of 20, 50, 80, 120, 200, 400, and 750 nm diameter were studied at three different peak intensities of $8.8 \times 10^{12}$, $1.3 \times 10^{13}$, and $1.8 \times 10^{13}$ W/cm$^2$. Figure 3 shows a representative selection of non-inverted 2D integrated VMI images from 20, 120, and 400 nm particles at these three peak intensities. Each row shows how the photoelectron spectra changed with increasing laser intensity but for a fixed particle diameter. At larger intensities, the electron momentum distribution extends to larger values (i.e. the cutoff energy increases) while the emission pattern becomes somewhat elongated along the polarization direction. Following a specific column in figure 3 shows how the emission pattern depends on the particle size at a fixed laser intensity. An increase in the particle diameter leads to more energetic electrons along with different angular distributions.

An obvious asymmetry in the photoelectron distribution can be seen along the propagation direction, especially for the 400 nm nanoparticles (bottom row of figure 3). This has been previously observed in Ref. [19] where, for few-cycle pulses, the carrier-envelope phase dependent part of the most energetic electron emission showed direct mapping of the induced near-field of the nanoparticle. Here, for $\rho \geq 1$, contributions from higher-order multipoles result in the asymmetry. These propagation effects are also referred to as nano-focusing and result in stronger field localization towards the propagation axis and larger field enhancement [34].

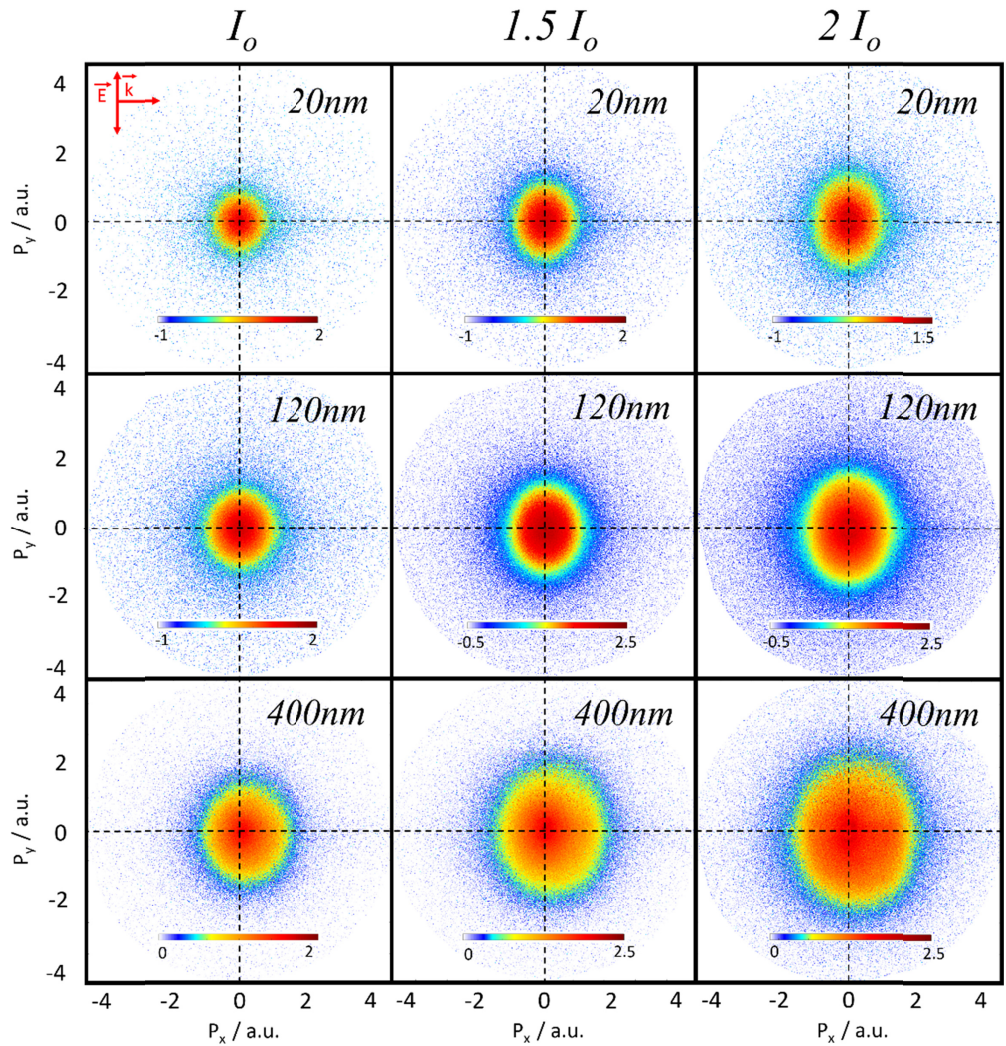

Fig. 3. Collection of integrated VMI images (rescaled to atomic units (a.u.) of momentum) as a function of diameter and peak laser intensity. Horizontal rows show the images for a constant nanoparticle diameter whereas vertical columns correspond to a fixed laser intensity. Dotted cross-hairs are placed at center of VMI images. $I_o = 8.8 \times 10^{12}$ W/cm$^2$. Color bar in log scale.

Figure 4 plots the cutoff energy of the photoelectron spectra as a function of nanoparticle size and laser intensity. The overall trend indicates a monotonic increase of absolute cutoff energy for both increasing diameter and incident laser intensity. Remarkably, the largest nanoparticle (750 nm) emits electrons almost six times more energetic than the smallest (20 nm) nanoparticle, given the exact same incident laser intensity. Qualitatively, there are two main factors which are likely to contribute to this electron energy increase. The first one is the size dependence of the near-field enhancement. As shown in the inset of figure 4, the maximal increase in local intensity due to the near-field, which is the proportional to the square of the field, increases by a factor of ~2.5 over the range of particles sizes studied [35]. Therefore, even though the near-field enhancement significantly contributes to the size-dependence of the maximum photoelectron energy, it does not fully account for the magnitude of the observed increase. The second major contributing factor to this increased cutoff energy is the larger total number of emitted electrons for larger particles, which results in enhanced charge interactions [19].

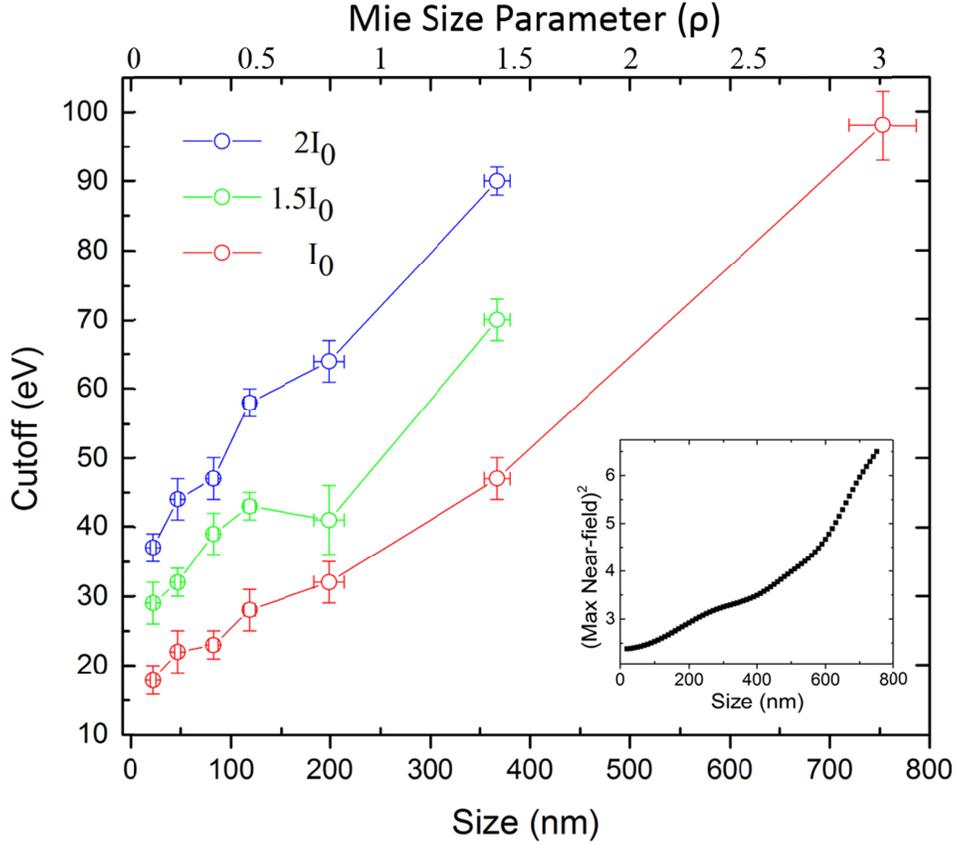

Fig. 4. Photoelectron cutoff energy as a function of SiO$_2$ nanoparticle size (bottom axis) and Mie size parameter (top axis) for three different peak laser intensities. Electron cutoff energy found using integrated non-inverted VMI images as seen in figure 3 and taking a radial distribution (scaled to energy units) as shown in figure 2(d). Particle diameters range from 20 - 750 nm. $I_o = 8.8 \times 10^{12}$ W/cm$^2$. Inset shows the squared maximum near-field intensity calculated by numerically solving the Mie equations as described in [35].

In order to reveal the effect of the incident light intensity, in figure 5 we plot the same results presented in figure 4 but with the cutoff photoelectron energies normalized by the ponderomotive energy, $U_p$, which scales linearly with intensity (open symbols). To facilitate the comparison with earlier data obtained at a different wavelength, in this graph we plot the results as a function of the Mie size parameter $\rho$, which is scaled to the wavelength. As can be seen from the figure, normalization in units of $U_p$ yields a nearly intensity-independent cutoff value in units of $U_p$ for each nanoparticle size. This indicates that, for the diameter and intensity range studied here, the electron cutoff energy scales linearly with peak driving laser intensity. This observation agrees well with the findings of earlier studies on SiO$_2$ particles with a comparable intensity range [19,36], as well as with the theoretical results using a quasi-classical mean-field Mie Monte Carlo (M$^3$C) simulations [19]. As can be seen from figure 5, where the experimental results from [19] are plotted as black squares, the cutoff energy values obtained in the present work agrees with the results for few-cycle pulses at small Mie parameters. However, our results for 25 fs pulses quickly diverge towards much larger cutoff energies for Mie size parameters larger than 1 than what was observed with few-cycle laser pulses.

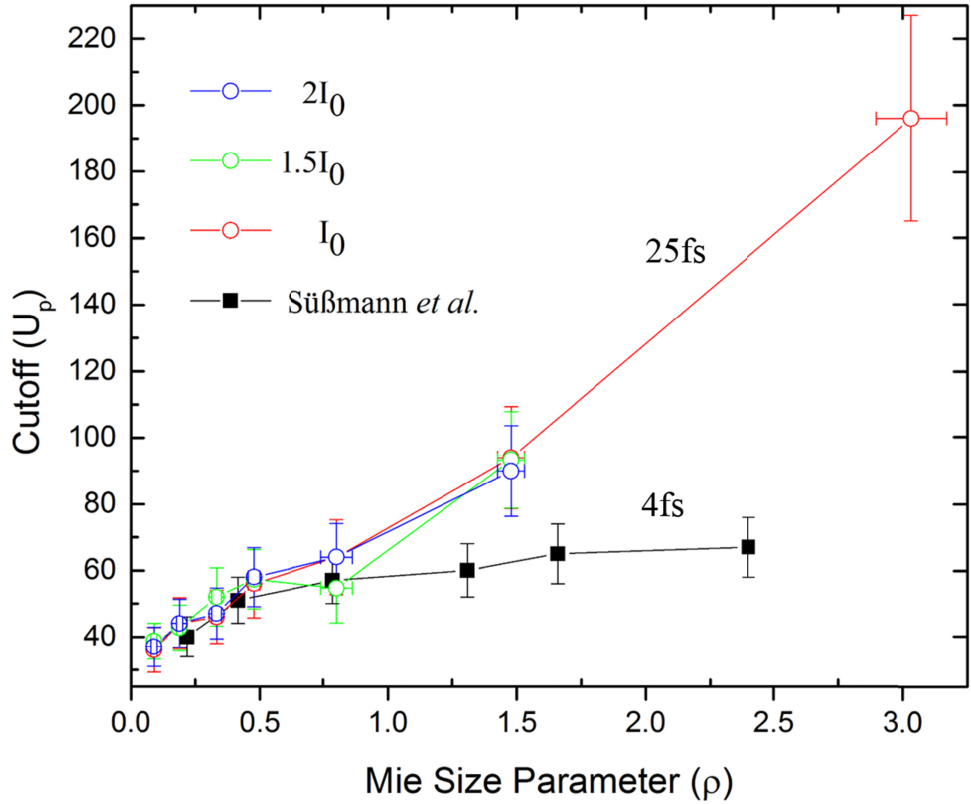

Fig. 5. Size dependent photoelectron cutoff energy values from $SiO_2$ nanoparticles normalized in units of the ponderomotive potential, $U_P$, as a function of the Mie size parameter. Open circles: electron cutoff values from the present work (25 fs pulses, central wavelength 780 nm) for three different intensities. $I_o = 8.8 \times 10^{12}$ W/cm$^2$. Black squares: data from Süßmann et al. [19] (4 fs pulses, central wavelength 720 nm, intensity $3 \times 10^{13}$ W/cm$^2$).

In the theoretical results of [19,24,27], the overall increase of cutoff energies with particle size was explained by linear near-field enhancement (cf. inset in figure 4), and nonlinear charge interactions. The latter consist of mainly two contributions: i) a trapping field forming near the surface mediated by residual ions, and assisted backscattering during the recollision phase and ii) Coulomb explosion of the emitted electron bunch. The number of created electrons with a longer laser pulse (25 fs in this study) is effectively larger than for a few-cycle pulse (4 fs employed in [19]) of the same peak intensity. Therefore, for longer pulses, a deeper trapping potential enhances the energy gain from the recollision process. Furthermore, an increased number of emitted electrons contributes to a rise in cutoff energies due to the repulsive forces within the electron bunch. While both processes (trapping field and Coulomb explosion) explain higher cutoffs for longer pulses, we have observed that the pulse duration dependence is not clearly distinguishable for Mie parameters smaller 1. A tentative explanation might be derived from previous theoretical work, where the contribution of these nonlinear acceleration processes was studied as a function of particle size [24] . It was found that the trapping field contribution (in units of $U_p$) remains roughly constant, while the Coulomb explosion contribution grows significantly for larger particles due to the stronger confinement of the initial electron bunch. We thus tentatively attribute the observed size dependence to the important role of field confinement and stronger repulsion between emitted electrons. A more definite and quantitative explanation of the observed pulse duration dependence, however, requires the challenging extension of the theoretical analysis performed in [24] to longer, multi-cycle pulses.

6. **Conclusion**

In this work, we have presented the results of the systematic study of laser-induced photoelectron emission from dielectric nanoparticles ($SiO_2$) as a function of particle size (ranging from 20 nm to 750 nm) and introduced an experimental technique which visualized the effects of focal volume averaging on the photoelectron spectra of gas-phase nanoscale targets.

Our approach is based on using the number of emitted photoelectrons per laser shot for a given nanoparticle size as a relative measure of the local laser intensity (i.e. the position of the particle within the laser focus). By sorting results according to this observable, accurate energy- and angle-resolved photoelectron spectra corresponding to a specific incident laser intensity range can be obtained. At the same time, we demonstrate that the maximal photoelectron energy (the photoelectron cutoff) can be accurately extracted from the focal-volume integrated data. While the cutoff energy increases nearly linearly with the peak laser intensity, it also strongly depends on the particle size, for a fixed intensity increasing by almost a factor of six over the range of particles diameters studied. Comparison of the size-dependent field enhancement factors and the observed increase in electron cutoff energies confirm that both the enhancement of the near-field and charge interactions are important for the formation of the high-energy part of the photoelectron spectrum.

Comparing our results obtained with a 25 fs pulse duration with earlier data for few-cycle (4 fs) pulses [19], we observe that, when scaled to the ponderomotive energy, $U_p$, both data sets agree well at small Mie size parameter ($\rho \ll 1$), whereas, within the studied intensity and Mie parameters, for increasing particle size ($\rho \geq 1$) the longer pulses produce much more energetic electrons. This difference correlates with stronger localization of the induced near-field on the nanoparticle due to propagation effects. In view of earlier theoretical analysis, our data suggest that this pulse duration dependence is due to an increase in charge creation and interaction with longer pulses, resulting in the additional energy gain [24]. Our findings demonstrate that the transition from few-cycle to multiple-cycle pulses can drastically change the response of nanoscale objects to an intense laser field. These results are expected to be relevant for the strong-field induced electron emission from other nano-scale targets with size-dependent near-field structure, in particular, plasmonic nanotips [11,12].

**Funding**

This work was supported by the Air Force Office of Scientific Research (AFOSR) grant FA9550-17-1-0369. S.-J.R. and A.M.S. were partially supported by Chemical Sciences, Geosciences, and Biosciences Division, Office of the Basic Energy Science, U.S. Department of Energy Award No. DE-FG02-86ER13491. The PULSAR laser was provided by Grant No. DE-FG02-09ER16115 from the same funding agency. A.M.S. acknowledges the support from Department of Defense (DoD) National Defense Science & Engineering Graduate Fellowship. P.R., J.S. and M.F.K. were supported by Max Planck Society; German Research Foundation (DFG) grant Kl-1439/8-2 (SPP1840) and Munich Centre for Advanced Photonics (MAP).

**Acknowledgements**

We thank Vinod Kumarappan, Varun Makhija and Aram Vajdi for providing the hit finding software for VMI data acquisition and analysis, and Yubaraj Malakar for help with the spectrometer design and commissioning.